\documentclass[aps,prl,reprint,superscriptaddress,showpacs]{revtex4-2}
\usepackage[utf8]{inputenc}
\usepackage[dvips]{graphicx}
\usepackage{epsfig}
\usepackage{multirow}
\usepackage{ulem}
\usepackage{textcomp}
\usepackage{url}
\usepackage[usenames,dvipsnames]{color}

\begin{document}

\title{Pseudospin-phonon pretransitional dynamics in lead halide hybrid perovskites}
\author{B. Hehlen}
\affiliation{Laboratoire Charles Coulomb, UMR 5221 CNRS-Universit\'e de Montpellier, FR-34095 Montpellier, France}
\author{P. Bourges}
\affiliation{Laboratoire L{\'e}on Brillouin, CEA-CNRS, Universit\'e Paris-Saclay, CEA Saclay, 91191 Gif-sur-Yvette, France}
\author{B. Ruffl\'e}
\affiliation{Laboratoire Charles Coulomb, UMR 5221 CNRS-Universit\'e de Montpellier, FR-34095 Montpellier, France}
\author{S. Cl\'ement}
\affiliation{Laboratoire Charles Coulomb, UMR 5221 CNRS-Universit\'e de Montpellier,  FR-34095 Montpellier, France}
\author{R. Vialla}
\affiliation{Laboratoire Charles Coulomb, UMR 5221 CNRS-Universit\'e de Montpellier,  FR-34095 Montpellier, France}
\author{A.C. Ferreira}
\affiliation{Laboratoire L{\'e}on Brillouin, CEA-CNRS, Universit\'e Paris-Saclay, CEA Saclay, 91191 Gif-sur-Yvette, France}
\affiliation{Univ Rennes, INSA Rennes, CNRS, Institut FOTON - UMR-6082, F–35000 Rennes, France}
\author{C. Ecolivet}
\affiliation{Univ Rennes, CNRS, IPR (Institut de Physique de Rennes) - UMR-6251, F-35000 Rennes, France}
\author{S. Paofai}
\affiliation{Univ Rennes, CNRS, ISCR (Institut des Sciences Chimiques de Rennes) - UMR-6226, F-35000 Rennes, France}
\author{S. Cordier}
\affiliation{Univ Rennes, CNRS, ISCR (Institut des Sciences Chimiques de Rennes) - UMR-6226, F-35000 Rennes, France}
\author{C. Katan}
\affiliation{Univ Rennes, CNRS, ISCR (Institut des Sciences Chimiques de Rennes) - UMR-6226, F-35000 Rennes, France}
\author{A. L\'etoublon}
\affiliation{Univ Rennes, INSA Rennes, CNRS, Institut FOTON - UMR-6082, F–35000 Rennes, France}
\author{J. Even}
\affiliation{Univ Rennes, INSA Rennes, CNRS, Institut FOTON - UMR-6082, F–35000 Rennes, France}

\date{\today}

\begin{abstract}

\noindent The low frequency lattice vibrations and relaxations are investigated in single crystals of the  four 3D hybrid organolead perovskites, MAPbBr$_3$, FAPbBr$_3$, MAPbI$_3$, and $\alpha$-FAPbI$_3$, at the Brillouin zone center using Raman and Brillouin scattering and at the zone boundary using inelastic neutron scattering. 
The temperature dependence of the PbX$_6$ lattice modes in the four compounds can be renormalized into universal curves, highlighting a common vibrational dynamics at the cubic to tetragonal transition. 
In particular, no soft vibration is observed excluding a displacive-like  transitional dynamics. 
The reorientational (pseudospin) motions of the molecular cations exhibit a seemingly order-disorder character recalling that of plastic crystals, but attributed to a secondary order-parameter. 
At ultra-low frequency, a quasi-elastic component evidenced by Brillouin scattering and associated to the unresolved central peak observed in neutron scattering, is attributed to center of mass anharmonic motions and rattling of the molecular cations in the perovskite cavities. 
Its partially unexpressed critical behavior at the transition
points toward the general importance of defects in HOPs preventing the  net divergence of order parameter correlations at the critical temperatures.

\end{abstract}

\maketitle
\section{Introduction}

A large variety of cristalline materials with outstanding fundamental and technological interests adopt the perovskite structure of general formula ABX$_3$. 
Simple oxide-based compounds (X=O) have early been considered as model systems of displacive phase transitions, i.e. controlled by a low-lying soft lattice vibration,  \cite{Lea70,Gla72,Fuj74,Kaw02,Cow11,Ben13}, the archetypal example being SrTiO$_3$   \cite{Sha72}.
However, with the onset of criticality, the growth of correlations drives a crossover from a weakly anharmonic (displacive) regime to an anharmonic (order-disorder) regime, where typically the phase transition is controlled by the freezing of a relaxational motion.
This behavior is amplified in the mixed perovskites, such as the ferroelectric relaxors of general formula (AA')BO$_3$ and A(BB')O$_3$, where a strong atomic, positional, and nanoscale polar disorder comes into play \cite{Cow11,Heh16}. 
In all the above-mentioned materials, the slowing down of the atomic dynamics, either vibrational or relaxational,  takes the form of power laws, ($T-T_c$)$^\beta$, where $T_c$ is the transition temperature and $\beta$ is a critical exponent \cite{Sco74}.
The value of the exponents characterizes the  {\it dynamical} critical  phenomena. They inform on the nature of the medium and long range correlations rather than on the structural details at atomic scale, and are therefore profound markers on the way a system evolves macroscopically from one thermodynamic equilibrium to another upon temperature.

Hybrid organolead materials (HOP) constitute another class of  perovskites, with lead (Pb) on the B-site, halide (Cl, Br, I), on the X-site, and an organic molecule (methylammonium, CH$_3$NH$_3^+$,  MA, or formamidinium, CH(NH$_2$)$_2^+$ FA)  on the A-site.  
They have established themselves in the field of photovoltaics, mainly owing to their optoelectronic properties and solar cell efficiciency. 
From a structural point of view, MAPbBr$_3$ (MAPB), MAPbI$_3$ (MAPI), FAPbBr$_3$ (FAPB), and FAPbI$_3$ (FAPI) exhibit a  similar series of phase transition with temperature.  
They are cubic ($Pm\bar 3 m$) at high temperature (except FAPI for which a possible trigonal phase has been also discussed at room temperature\cite{Bin15}), tetragonal at intermediate temperatures, and then orthorhombic (Pnma) at low temperature \cite{Ono90,Swa03,Whi16,Sch17,Fab16}.
The tetragonal space group is $I4/mcm$ except in FAPB where it is $P4/mbm$~\cite{Sch17}.
In MAPB, an intermediate phase between the tetragonal and orthorhombic phase has also been reported \cite{Ono90}.
These transitions mostly relate to the tilting of the PbX$_6$ octahedra. 
Similarly to the model compounds SrTiO$_3$\cite{Lea70,Gla72} and CsPbCl$_3$\cite{Fuj74}, where the staggered rotation angle is the primary order parameter of the phase transition, the antiphase tilting along one fourfold axis at the cubic to tetragonal transition has been emphasized \cite{Bee16}. 

The lattice dynamics of HOPs have been investigated by numerical simulations \cite{Bri15,Per15,Pon18,Pon18b,Pon19}, inelastic neutron scattering (INS) \cite{Swa15,Dru16,Let16,Fer18,Li17,Son19,Fer20,Zha20,Wea20}, inelastic X-ray scattering \cite{Bee16,Com16}, infrared absorption \cite{Bak15} and Raman scattering (RS) \cite{Bri15,Qua14,Par17,Yaf17}, to cite the most conventional techniques.  
However, very few experiments focus on  the vibrational modes  at frequency below 100 cm$^{-1}$ ($\sim$ 12 meV)  where lie the vibrations of the inorganic PbX$_6$ sublattice \cite{Yaf17,Fer20, Par17}.  
One limitation is likely the strongly anharmonic behavior of these materials  at high temperature which complicates the data analysis \cite{Fer20}.   
Indeed, the orientational disorder of the organic molecules distort the inorganic scaffold and in addition, whether thermally excited \cite{Che15} or laser-induced by transient absorption \cite{Par18}, the relaxational motions couple with vibrations of the PbX$_6$ octahedra.
This induces homogeneous and possibly inhomogeneous line broadening and overlapping, yielding to the complex phonon responses observed experimentally in both RS and INS \cite{Fer20,Zha20}.
Another intrinsic difficulty for the analysis of HOPs  by comparison to other crystals, is the presence of several low frequency phonon bundles~\cite{Let16}. 
As a consequence, it is still not clear whether the long range ordering of the PbX$_6$ sublattice is induced by the slowing down of a low-frequency soft vibration. 

The role of the relaxational motions at the transitions is not clear either.  
An analysis based on calorimetric measurements has long ago predicted an order-disorder transitional mechanism in MAPB due to the orientational ordering of the MA molecule into its cuboctahedral environment \cite{Ono90}. 
Such orientational ordering has latter been confirmed \cite{Che15,Leg15} and possibly combines with translational degrees of freedom either connected to optical or acoustic phonon modes \cite{Yaf17,Let16}. 
At high temperature the organic cations are moving almost freely in a cubic potential and with little interactions with the PbX$_6$ cages \cite{Eve16}.
Going down to lower temperature, 
the reordering of the PbX$_6$ sublattice to lower symmetry structures imposes steric effects and freezes the molecular relaxational degrees of freedom.

These experimental observations recall phase transitions in molecular crystals and in plastic crystals, namely from a molecular disordered phase at high temperature to a perfectly ordered crystal at low-temperature. In the disordered phase, the molecular motions can be rotational and translational but it is considered that for plastic crystals, the molecular centers of mass remain located on the average at well-defined crystallographic positions.  If the dynamical disorder shows off as stochastic molecular reorientations rather than conformational disorder, the phase is frequently referred to as a plastic crystal or rotor phase \cite{Lyn94}.
In molecular crystals, a pretransitionnal dynamical regime directly related to the coupling between organic molecules has a character ranging from  displacive to  order-disorder which can be interpreted within a coupled double-well  potential model \cite{Bou96,Bau76,Cai86}.
More, structural phase transitions in molecular crystals with a full or partial displacive character, may exhibit a quasi-elastic central peak, which has been tentatively attributed to defects or impurities in many inorganic systems \cite{Hal76}. 
The latter case of plastic crystals is characterized by a broad distribution of relaxational motions and follows a mechanism exclusively of order-disorder type typically analyzed with the pseudospin-phonon concept, where pseudospin motions relate to orientational degrees of freedom of molecules \cite{Lyn94, Yam74}. 
A displacive behavior of phonons can trigger the ordering of molecular pseudospins, when the  molecules undergo a fast relaxational motion. 
On the other hand in the case of a slow relaxation, the spectrum shall exhibit a triple peak structure with almost fixed phonon side peaks and an intrinsic central component showing a critical behavior.
Indeed a recent work in MAPI rather reports on the observation of a central peak attributed  to extrinsic mechanisms leading in turn to a first order character for the phase transition \cite{Wea20}.  

The motivation of this work is to provide a detailed analysis of the vibrational and relaxational critical phenomena at the cubic to tetragonal transition in HOPs, that is at temperatures close to working conditions of these materials. 
RS and INS will be used to probe relaxational motions and to look for a soft mode at Brillouin zone center and Brillouin zone boundary close to superstructure peaks, and BS will be used to investigate ultra-low frequency relaxations.

\section{Experimental details}
Experiments have been preformed in 
single crystals of MAPbBr$_3$ (MAPB), FAPbBr$_3$ (FAPB), MAPbI$_3$ (MAPI), and $\alpha$-FAPbI$_3$ (FAPI). 
The synthesis has been previously described in details in our reports on INS experiments \cite{Fer18,Fer20}. 
All single crystals were grown at  the Institut des Sciences Chimiques de Rennes (ISCR) using inverse temperature crystallisation following recipies given in Refs. \cite{Sai15,Zhu16}.\\ 

Raman scattering has been performed in the back scattering geometry using an optical microscope with a $\times$\,100 objective. 
Two ultra sharp Notch filters \cite{Bragg} were placed on the scattered light trajectory to filter the strong elastic component. 
The Raman spectra were collected using a single grating spectrometer with 1800 groove/mm.    
This setup allows measurements down to frequencies $\pm$~10~cm$^{-1}$ typically. 
In MAPB, FAPB, and MAPI, the wave vector {\bf q} of the incident radiation was along the [001] crystallographic direction and its polarization was parallel to the [100] or [110] axes.
For each of these two  incoming geometries, the scattered beam was analysed  parallel (polarized spectra, $I_\parallel$) and perpendicular (depolarized spectra, $I_\perp$) to the incident  polarization, leading therefore to four spectra per sample.   
To avoid laser-induced degradation,  below bang-gap excitation where used in MAPB ($\lambda$~=~660~nm and 852~nm), FAPB ($\lambda$~=~852~nm), and MAPI ($\lambda$~=~852nm). 
In FAPI we used an excitation wavelength $\lambda$~=~532~nm, that is above the band-gap,  but with a very low laser power of  $\sim$~400~$\mu$W.

Following our previous reports \cite{Fer18,Fer20}, INS measurements were conducted on the cold neutron (4F2) and on the thermal (1T) triple axis spectrometers (TAS), all located at the Orphée reactor (CEA Saclay) working in the constant final neutron energy mode. For 4F2, the final neutron wavevector was equal to $k_F$=1.55 \AA$^{-1}$ with the use of a Be filter on the $k_F$ arm to remove higher harmonics. For 1T, the final neutron wavevector was kept constant to $k_F$=2.662 \AA$^{-1}$ with the use of a pyrolytic graphite  filter on the $k_F$ arm to remove higher harmonics. Constant-{\bf Q} scans were perfomed at the R- and M-points of the  Brillouin zone-boundary.
These points correspond to the momentum space where superstructure peaks occur in the lower temperature phase similarly to what was already discussed in CsPbCl$_3$ \cite{Fuj74}, and more recently in CsPbBr$_3$ \cite{Lan21}, where two-dimensional overdamped fluctuations were observed along the R-M line in the high temperature phase. 

The data have been complemented by Brillouin scattering  experiments for probing  ultra-low frequency relaxations. 
Spectra were recorded at 647.1~nm in a pseudo back-scattering geometry, by using a krypton ion laser and a tandem of Fabry-Perot interferometers where each interferometer is triple passed giving a contrast larger than 10$^{12}$.
This setup prevents for multiple order overlapping, and is therefore well suited for measuring quasi-elastic spectral components.    

\section{Results}
\subsection{MAPB from low to high temperature} 
\begin{figure}[h]

\includegraphics[width=8.5cm]{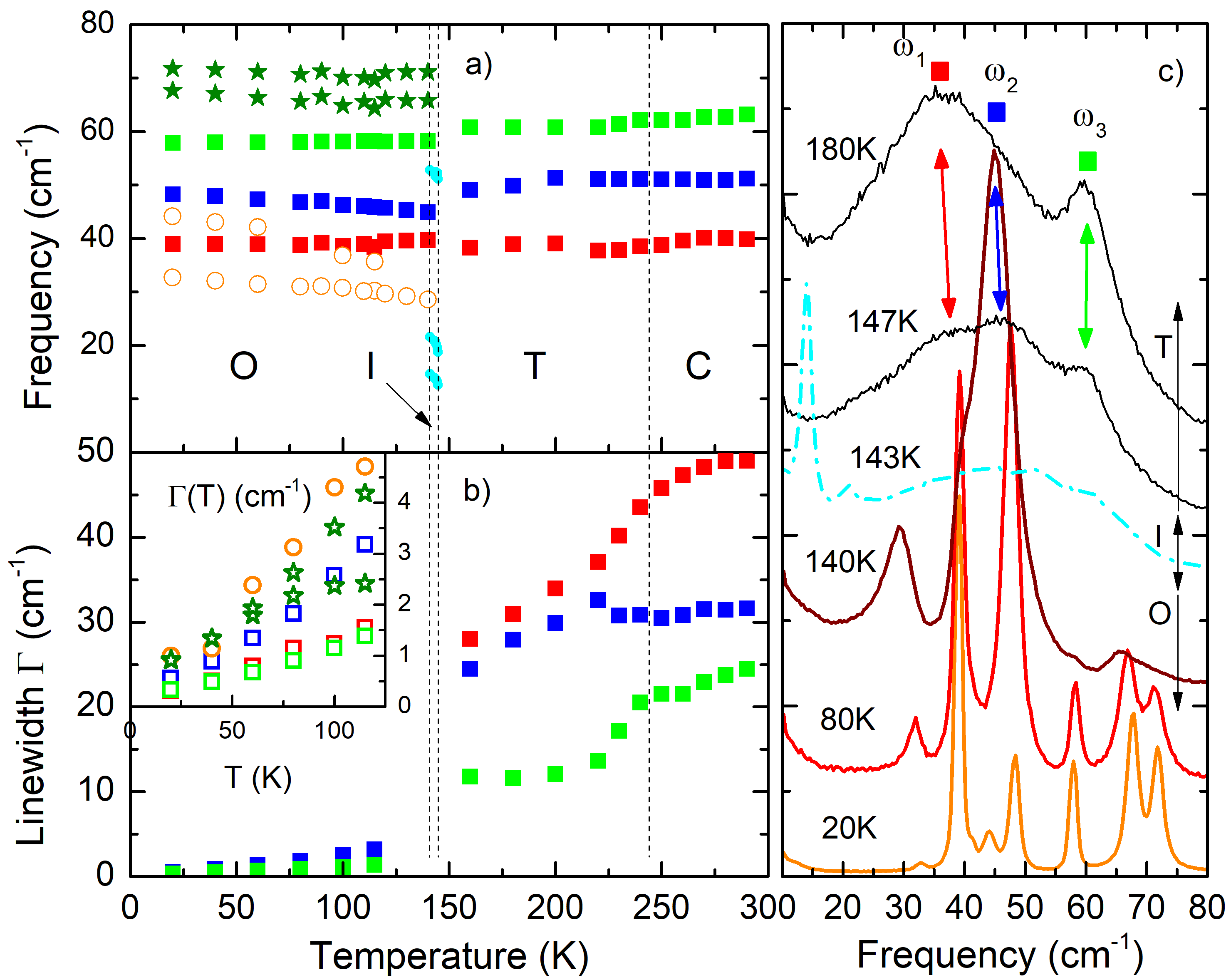}
\caption{Raman spectroscopy in MAPB across the three phase transitions : cubic (C), tetragonal (T), Intermediate (I), and orthorhombic (O). 
{\bf a)} phonon frequency and {\bf b)} phonon linewidth (zoom in the inset) resulting from a fit with damped harmonic oscillators. 
{\bf c)} Low-frequency Raman responses in the tetra (T), intermediate (I), and orthorhombic phase (O).
The main phonon bands $\omega_1$, $\omega_2$, and $\omega_3$, are identified by red, blue, and green filled squares (panels a,b,c) and colored double arrows (panel c).} 
\label{fig:RSLowT}
\end{figure}

In order to ensure the coherency of the Raman data treatment of the vibrational spectra, we measured  the optical phonons of MAPB from the orthorhombic phase where they are well defined up-to the cubic phase where they become highly damped. 
We identified eight bands below 80\,cm$^{-1}$ in the orthorhombic phase (Ref. \cite{Fer20} and Fig. 1) .   
Among them, three main responses $\omega_1$, $\omega_2$, and $\omega_3$, can be followed up to high temperature as shown from a direct inspection of the raw data  in Figure 1c. 
They are identified by squared symbols and are claimed to be related to rocking and bending of the inorganic halide octahedra \cite{Swa15,Fer20}. 
Their frequency have an almost flat temperature dependence (Fig. 1a) while the linewidth are all strongly affected by the structural instabilities (Fig. 1b).
They initiate a smooth narrowing close to the cubic to tetragonal transition and  an abrupt one at the tetragonal to orthorhombic transition. 
This line narrowing provides an indirect way to observe the freezing of the relaxational motions of the MA molecule through their coupling with lattice vibrations.
At the onset of the tetragonal phase, the reorientational motions of the C-N axis of the MA molecule between  different crystallographic axes progressively freeze on cooling, and at the first order tetragonal to orthorhombic transition they are all frozen \cite{Che15}, leading to the steep narrowing of the Raman bands.  
For temperatures below it only remains coupling with rotations of the MA molecule around the C-N axis, whose reduced strength on cooling likely translates into the line narrowing shown in the caption of Fig. 1b.

Three additional modes at $\simeq$~13~cm$^{-1}$, $\simeq$~20~cm$^{-1}$, and $\simeq$~52~cm$^{-1}$ appear in the Raman spectra between$\sim$\,141\,K and $\sim$\,147\,K, confirming the existence of the intermediate phase (I) observed by calorimetry measurements \cite{Ono90}. 
The two modes at low frequency are very sharp as compared to the otherwise broad  Raman response, as shown by the spectra at 143~K in Fig. 1c. 
Their frequencies  are in addition below the first optical branch in the tetragonal phase. 
This might suggest that the I phase corresponds to an incommensurate phase and the two sharp modes would originate from acoustic phonon branches after folding of the Brillouin zone at the T-I transition. 
The fact that the broad Raman signal underneath is similar in the tetragonal and intermediate phases likely underpin the non-dispersive nature of the optical branches \cite{Fer20}.  
We also observed an hystereris effect on cooling and heating, and therefore the temperature interval of the phase I may slightly shift downward or upward depending on the thermal route.

Finally, no soft mode is evidenced at the tetragonal to orthorhombic phase transition suggesting that the latter is likely not displacive. 
If it was the case, one would observe in the low-temperature phase, the hardening of the zone boundary soft phonon of the tetragonal phase. This conclusion  contrasts with previous interpretation of  inelastic neutron scattering data where a  mode softening was speculated at the zone boundary in a powder  sample of MAPB \cite{Swa15}.

\subsection{Zone-center phonons: Raman scattering} 
\begin{figure}[t]
\includegraphics[width=8.5cm]{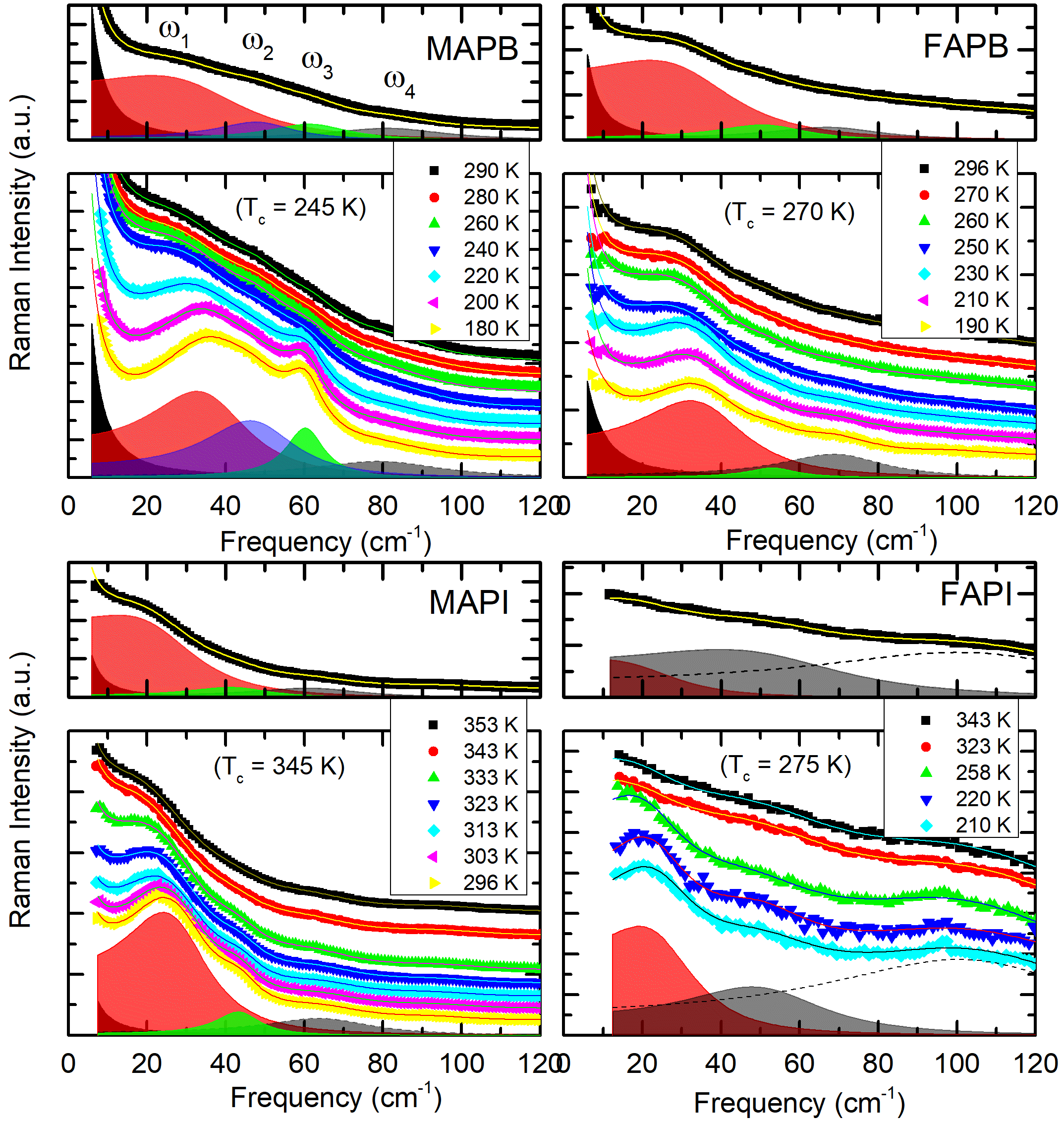}

\caption{Examples of Raman spectra (symbols) of various HOPs around thier respective  tetragonal to cubic phase transitions with  fits comprising overdamped harmonic oscillators and a central component (plain lines). 
For each compound, the individual spectral line shapes $S_i(\omega)$ ($i=1$ in red, $i=2$ in blue, $i=3$ in green, $i=4$ in gray) are shown for the lowest temperature spectra (bottom, tetragonal phase) and for the highest temperature spectra (top, cubic phase). For clarity, the spectra have been shifted up by a constant value.} 
\label{fig:RSspefit}
\end{figure}

A selection of Raman spectra obtained in the four hybrid compounds and collected across their respective cubic to tetragonal transition is shown in  Fig.\,\ref{fig:RSspefit}. 
The spectra presented in Fig.\,\ref{fig:RSspefit} corresponds to polarized spectra ($I_{\parallel}$) with incident polarization parallel to [110] and $q$ // [001].   
It is the situation where the low frequency mode  offers the most favorable contrast. 
They have been recorded at the same position into the sample in order to avoid modifications due to the formation of structural domains below $T_c$.\\
A direct observation already suggests that the responses in the cubic phase very much look like a smeared-out version of that of the tetragonal phase.
For example, in MAPB where $\omega_1$, $\omega_2$,  and $\omega_3$ are almost spectrally resolved at low temperature,  the modes strongly broaden with increasing temperature and finally lead to the broad and convoluted Raman response in the cubic phase. 
The behavior is very similar in FAPB and MAPI in which  $\omega_2$ is absent or very weak for that scattering geometry.
The spectroscopy in FAPI is rather challenging and the spectra are limited to one single set obtained along an arbitrary cristallographic direction and performed without polarization analysis.  
Luckily, $\omega_1$ clearly develops in the tetragonal phase and could be followed with temperature. 
At high temperature we had to satisfy with the spectra of Fig.\,\ref{fig:RSspefit} showing an  almost monotonous intensity decay of the Raman signal with frequency. 

The spectra have been fitted using damped harmonic oscillators (DHO) accouting for $\omega_{1}$, $\omega_{2}$, $\omega_{3}$.  
The scattering spectral function for a given mode $i$ reads 
\begin{equation}
S_i(\omega)=A_i [\text{n}(\omega)+1]\times\text{Im}[1/(\omega_{0i}^2-\omega^2-i\Gamma_i\omega)]
\label{DHO}
\end{equation}
n$(\omega)$ is the Bose-Einstein population factor while $\omega_{0i}$, $\Gamma_i$ and $A_i$ are respectively the DHO frequency, damping and the amplitude of the mode submitted to a driving force (the incident electric field).
A fourth DHO at frequency slightly above the first three was required to ensure a good fitting quality ($\omega_4$). 
The analysis also required a quasi-elastic component (QE) (in black in Fig.\,\ref{fig:RSspefit}) of spectral form as,
\begin{equation}
S_{QE}(\omega)=A_{QE} [\text{n}(\omega)+1]\times{\rm Im}[i\omega/(\gamma_{QE}-i\omega)]
\label{QE}
\end{equation}
 where $\gamma_{QE}$ is the half width at half maximum and $A_{QE}$ the amplitude of the QE component. An accurate characterization of this component will be provided below by using high resolution Raman spectroscopy.

\begin{figure}[t]
\includegraphics[width=8cm]{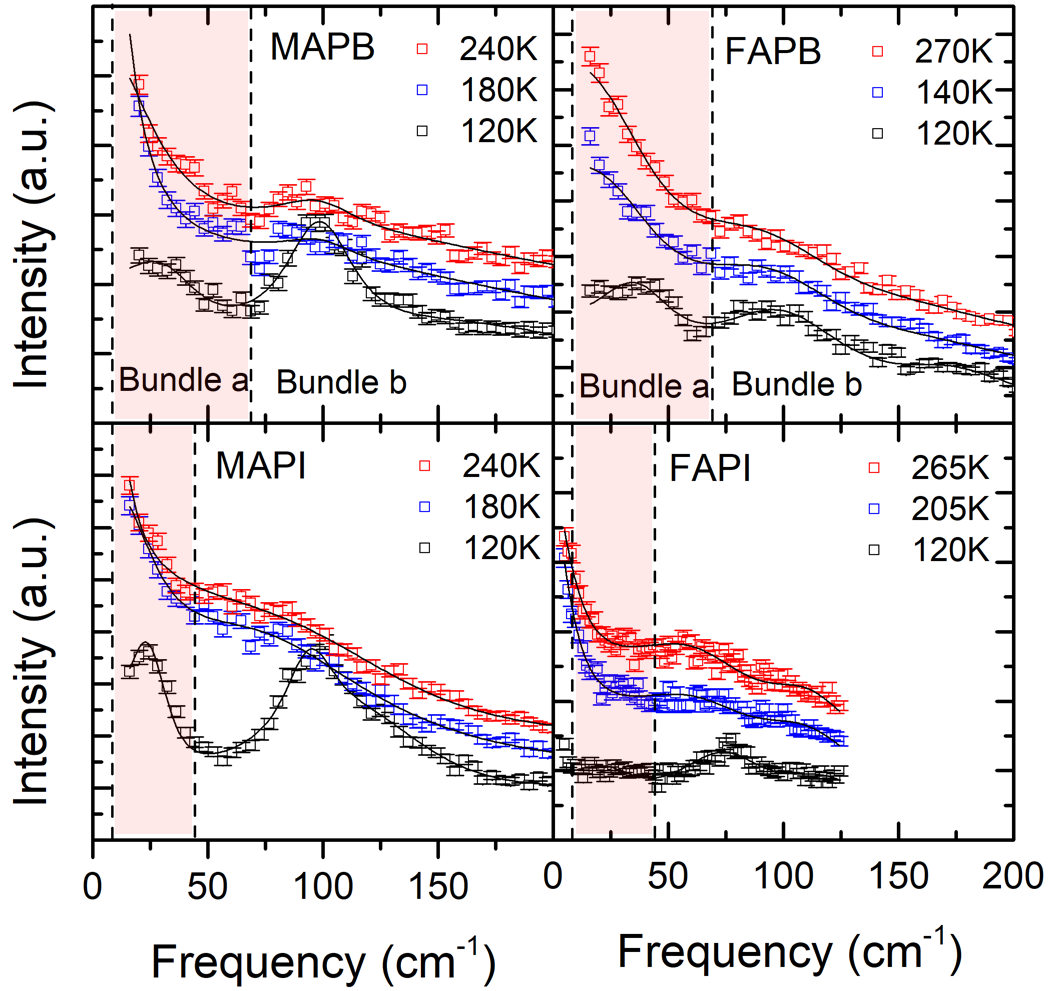}
\caption{Inelastic neutron scattering constant-{\bf Q} scans around the Brillouin zone boundary at the M-point {\bf Q}=(5/2 1/2 0) (FAPB) and the R-points {\bf Q}=(5/2 1/2 1/2) (MAPI) and (3/2 3/2 1/2) (MAPB and FAPI).
All experiments done on the thermal instrument (1T) except for FAPI measured on cold TAS (4F2). For each compound, superstructure Bragg peaks occur in the tetragonal phase at these {\bf Q}-positions. The lines are fits as explained in the text. The spectra at the two highest temperatures have been shifted up by a constant value of 50 and 100 counts.}
\label{fig:INS}
\end{figure}
In order to extract the individual spectral line shapes shown in Fig.\,\ref{fig:RSspefit}, we fitted the data collected in the four scattering geometries. 
Different Raman polarization geometries have been utilized to disentangle the phonons. 
The fact that the modes are more or less active depending on the scattering geometry provided a clear asset for deconvoluting the bands when they are strongly overlapping due to their large dampings, in particular at the onset of the cubic phase. 
For example, it is easier to follow $\omega_3$ from  the depolarized spectrum of MAPB for which $\omega_1$ is significantly weaker (Fig. S1 in the supplemental material \cite{SupMat}). 
It is the fitting of the complete set of data which give consistency to the fits of Fig.\,\ref{fig:RSspefit}, which otherwise would have been hazardous. 

\subsection{Zone-boundary phonons: inelastic neutron scattering}

INS has been performed at the Brillouin-zone boundary at momentum positions where the superstructure Bragg peaks occurs in the tetragonal and orthorhombic phases. 
As a result, INS experiments report on low energy optical phonons at the R-point in MAPI, FAPI and MAPB and at the M-point in FAPB (Fig. \ref{fig:INS}). 
The experimental conditions are those described in Ref. \cite{Fer20} and we will focus  first on vibrations typically below 60 cm$^{-1}$ ($\sim$ 7.5 meV),  which are primarily associated with  
the relaxational and vibrational properties of the PbX$_6$ cage. 
That spectral range corresponds to the phonon bundles labelled (a) in  Ref. \cite{Fer20}. It is highlighted by the shadowed region in Fig.~\ref{fig:INS}.
Bundle (b) at intermediate frequencies in Ref. \cite{Fer20} is also shown.
The broad feature centered around  40~cm$^{-1}$ (5~meV) in bromide samples and around 20-25~cm$^{-1}$ (2.5-3~meV) in iodide samples is attributed to rocking and bending of PbX$_6$ octahedra \cite{Fer20}.
It mixes at least the three lowest Raman modes $\omega_1$, $\omega_2$, $\omega_3$, not resolved in INS. 
Accordingly the bundle has been fitted as a whole using a single DHO (Eq. \ref{DHO}). 
Its frequency was fixed to the value at low-temperature, as suggested by the flat temperature-dependence of the Raman frequencies (Fig. 1a above and section ''Analysis and discussion'' below), and the fits were made without any quasi-elastic contribution. 
Despite these constraints, the results reproduces very well the experimental data and therefore confirm the initial hypothesis that Raman and neutron responses behave similarly.  
The apparent overdamped responses in the tetragonal and cubic phase (INS spectra in the cubic phase not shown but very similar to the highest temperature ones in Fig.~\ref{fig:INS}) arise from the progressive broadening of the three modes underlying bundle (a) rather than by a temperature softening of a vibration. 
However, it should be stressed that the modes remain underdamped with $\omega_{0}^2 > \Gamma^2/2$ (see below).

One should put our finding in perspective with the litterature which generally attributes the broadening of the low energy phonon branches to a soft phonon. In MAPB, the previous INS data \cite{Swa15} have been measured on a powder sample. 
The supposedly phonon softening at the tetragonal-orthorhombic transition is deduced from  Q-integrated and powder-averaged data where the information on the momentum space is lost. 
The overall thermal broadening of all phonons is misleadingly mimicking the softening behavior of a single mode.  In constrast, our inelastic neutron experiments clearly indicate no soft phonon at the phase transition. 
It should be stressed that these measurements directly probe in single crystals the phonon spectral response at M- or R- points in the tetragonal phase (Fig.~\ref{fig:INS}) and in the cubic phases above the transition temperature (not shown), where phonon softening is to be expected in a displacive-type phase transition \cite{Sha72,Sco74,Bou96} and where the critical fluctuations associated to the phase transition should occur. The absence of soft mode at the tetragonal-cubic transition was also reported in the chlorine materials MAPbCl$_3$ \cite{Son19}.
Phonons calculation using first principle methods often gives unstable modes over a large part of the Brillouin zone but it has recently been shown that these instabilities disappear under certain conditions~\cite{Pon18}, such as when calculations account properly for anharmonic effects~\cite{Pat15}.

\subsection{Insight into the quasi-elastic component}

\begin{figure}[h]
\includegraphics[width=7.15cm]{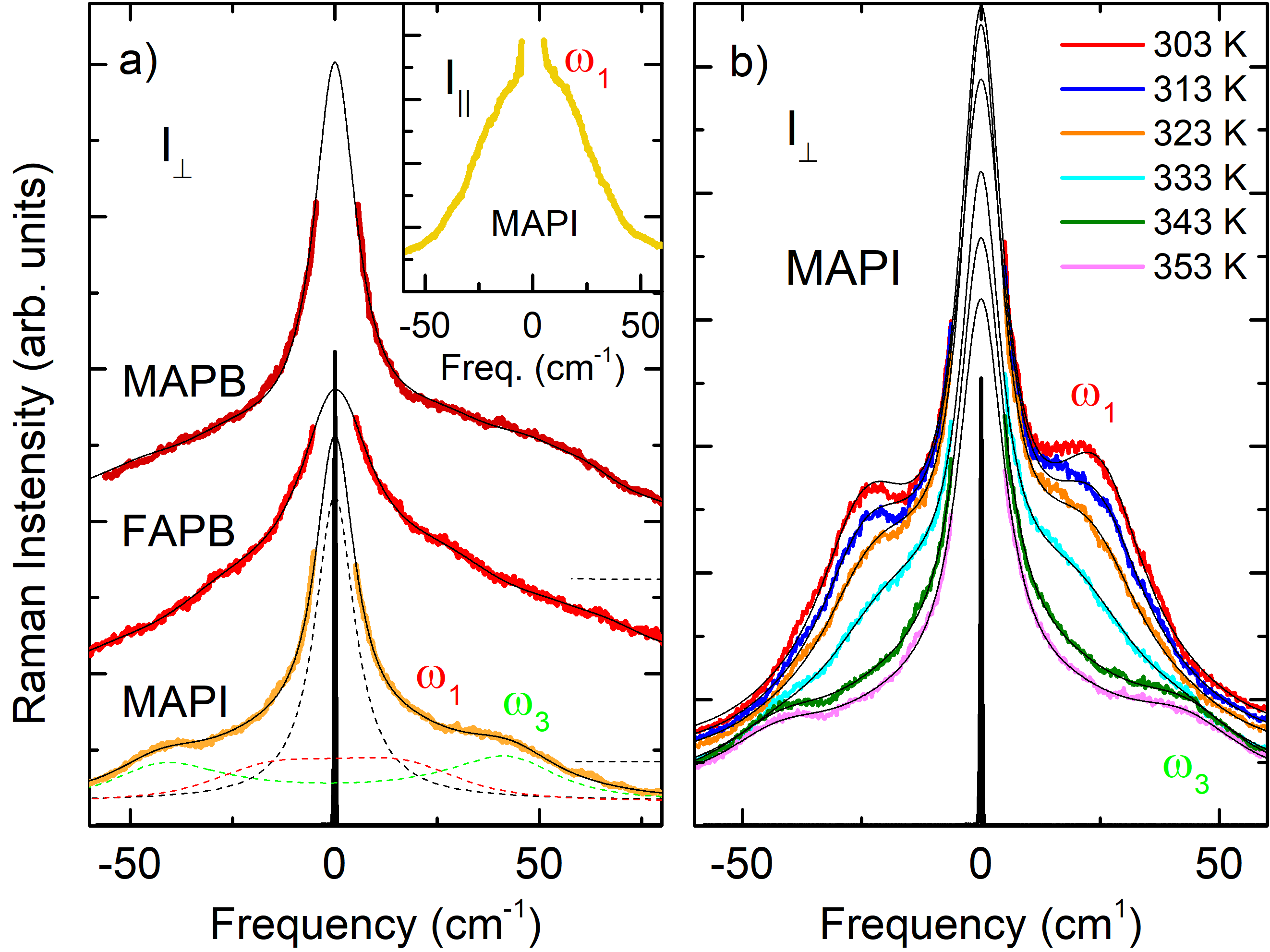}\hspace{.25cm}
\caption{{\bf a)} Depolarized Raman spectra ($I_{\perp}$) with incident polarization parallel to [100] and $q$ $\parallel$ [001] of MAPB, FAPB, and MAPI in the cubic phase. The inset shows the polarized spectra  ($I_{//}$) of MAPI. The dashed line indicates approximately the position of the peak maximum of $\omega_1$. {\bf b)} Temperature dependence of the low frequency Raman response in MAPI (I$_{\perp}$). 
}
\label{fig:QERS}
\end{figure}

A grating with 2400 grooves/mm was used to access very low frequencies in Raman scattering experiment (Fig.~\ref{fig:QERS}). 
In addition, the raw data were corrected by the transmission function of the Notch filters measured separately using a white light source. 
This setting allows measurements down to 4-5~cm$^{-1}$ with a spectral sampling of 0.16~cm$^{-1}$ per pixel on the CCD,  and provides an optimum situation for the measurement of quasi-elastic (QE)  components with linewidth as low as 
$\simeq$~2 cm$^{-1}$.   The results in the cubic phase of MAPB, MAPI, and FAPB are shown in Fig.~\ref{fig:QERS}a. 

One clearly observes a QE growing  below 10~cm$^{-1}$ in the depolarized spectra ($I_\perp$) with the incident polarization parallel to [100].
For this scattering geometry the mode $\omega_1$ is the weakest.  
By comparison, in the polarized spectra  ($I_{||}$) the central mode is still present but embedded into the strong response of $\omega_1$ (inset of Fig. \ref{fig:QERS}a).
The fits displayed in Figure\,\,\ref{fig:QERS} have been performed by fixing the DHO widths and frequencies to the values obtained from the spectra of the preceding section. 
Only their amplitude were allowed to vary, as well as the amplitude and the width $\gamma_{QE}^{RS}$ of the quasi-elastic component.  
Despite this very tight fitting constraint, the agreement with experiment is very good.  
At room temperature, we find  $\gamma_{QE}^{RS} \simeq$ 6.5\,cm$^{-1}$,  5.5\,cm$^{-1}$,  and 8.5\,cm$^{-1}$ in MAPB, MAPI,  and FAPB, respectively.  
It is worth noting that a model assuming a central mode with a broader linewidth, as proposed in Ref. \cite{Yaf17}, does not give reliable fits in any of the samples and for any scattering geometries explored.  
Conversely, relaxational motions with larger relaxation times (shorter linewidths) may of course exist, but are not resolved by our instrument. 
Figure \ref{fig:QERS}b shows the temperature dependence of the QE component of MAPI. 
Similarly to the low-resolution data in Fig. 2, one observes the  narrowing of $\omega_1$ when the sample is cooled down into the tetragonal phase, but on the other hand, the quasi-elastic component remains almost unchanged.  
In particular, its linewidth evolves very little with  temperature, at least in the resolution limit of our experiment ($\pm 1$~cm$^{-1}$), emphasizing a likely non-critical behavior.

\begin{figure}[h]

\includegraphics[width=7.2cm]{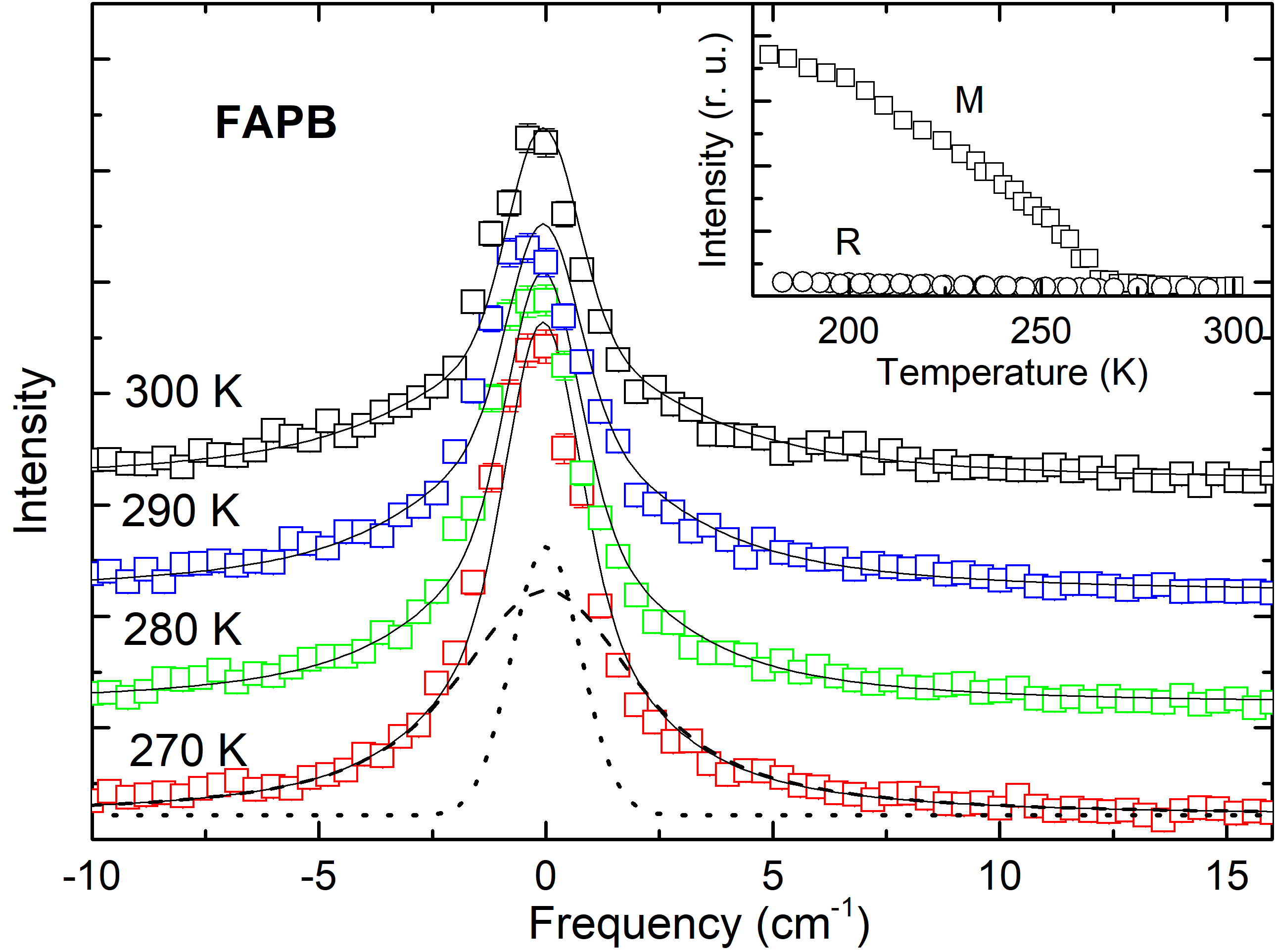} \\ \vspace{.5cm}
\caption{
Low-frequency INS response in FAPB at the  M-point {\bf Q}=(3/2 1/2 0). {\bf a)} Spectra at different temperatures  fitted with the sum a Lorenztian  curve given by Eq. \ref{QE} accounting for the QE, and a delta function (CP) accounting for the elastic-like central peak, convoluted by the instrument resolution function (dashed and dotted lines at 270 K, respectively). 
The spectra have been shifted by 200 counts for clarity.
The inset shows the temperature dependence of the elastic intensity at the M and R points. 
}
\label{fig:QEINS1}
\end{figure}

 In FAPB, we performed a detailled INS study of the low energy excitations at the M-Point at temperatures just above the tetragonal to cubic phase transition. 
 The latter occurs around $\sim$ 265 K as observed from the temperature dependence of the superstructure Bragg peak at {\bf Q}=(3/2 1/2 0) (inset of Fig.~\ref{fig:QEINS1}.)
 Figure \ref{fig:QEINS1} shows quasi-elastic scans at the same M-Point at temperatures T $>$ 265K. 
 The central intensity is slowly growing by cooling down from 300 K. 
 These spectra can be nicely fitted by the convolution product of the instrument resolution function by the sum of a Lorentzian function describing a QE response (INS QE) as Eq. \ref{QE}, in addition to a delta function.
 The latter around zero energy represents two possible contributions. 
 On the one hand, it primarily describes the elastic incoherent scattering which is generally not negligible in organic compounds \cite{Bou96}. 
 On the other hand, it can also correspond to a possible unresolved central peak (CP) which is typically observed at structural phase transition \cite{Sha72,Bou96}. 
 These responses stand on a constant background accounting for the  tail of the modes constituting bundle (a). 
 Its amplitude has been consistently imposed by the structure factor of the bundle (a) obtained from the fits of the scans measured with a wider energy range (Fig. \ref{fig:INS}).

\begin{figure}[h]
\includegraphics[width=8cm]{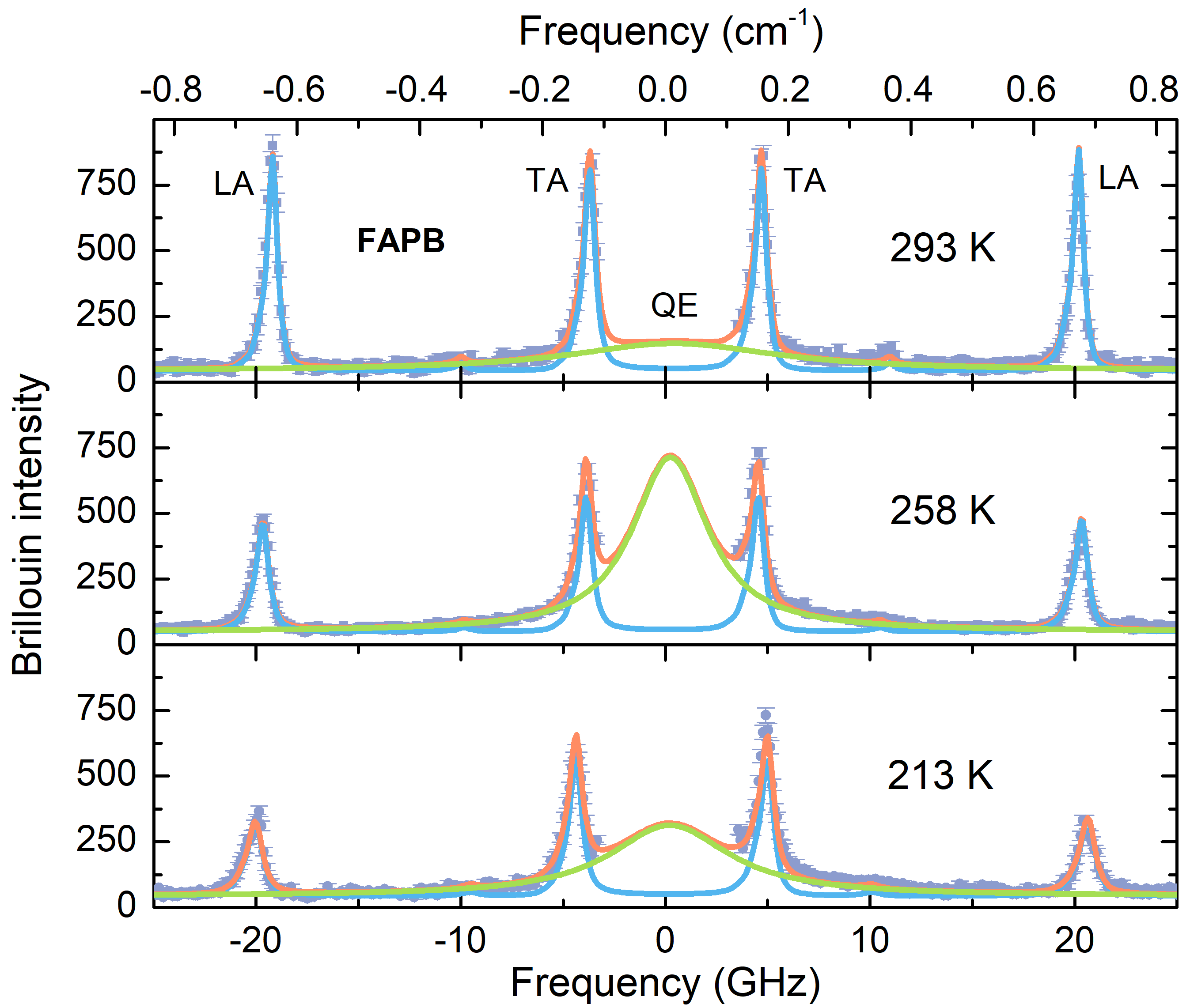}
\caption{Brillouin spectra in FAPB (symbols) and their fits (lines): transverse acoustic (TA) and longitudinal acoustic (LA) phonons (cyan), Quasi elastic (green), and the total (orange) which includes in addition a constant background accounting for the broad QE component (note that 30 GHz $\sim$ 1 cm$^{-1} \sim$ 0.12 meV) } 
\label{fig:BSspe}
\end{figure}

In the prospect of deciding whether the INS CP is purely elastic or quasi-elastic, Brillouin scattering (BS) is a technique of choice. 
Our previous experiments in MAPB~\cite{Let16} highlighted two quasi-elastic components. 
One broad and non-critical, of linewidth 30-100 GHz ($\sim$~1-3 cm$^{-1}$), namely in the energy range corresponding to the RS and INS QE mentioned above. 
In addition, a critical-like narrow component of linewidth $\sim$~10~GHz  ($\sim$~0.3~cm$^{-1}$) near T$_c$ was also observed. 
This promotes an interpretation of the INS CP in terms of a dynamical unresolved QE, as further discussed in the data analysis section below.
Indeed, a narrow component of linewidth $\sim$~5~GHz typically ($\sim$~0.15~cm$^{-1}$) is also observed in FAPB, as shown in Fig.~\ref{fig:BSspe}.
Its intensity reaches a maximum near T$_c$~=~265~K, suggesting in addition a critical-like behavior. 
\begin{figure*}[t]
\includegraphics[width=9.5cm]{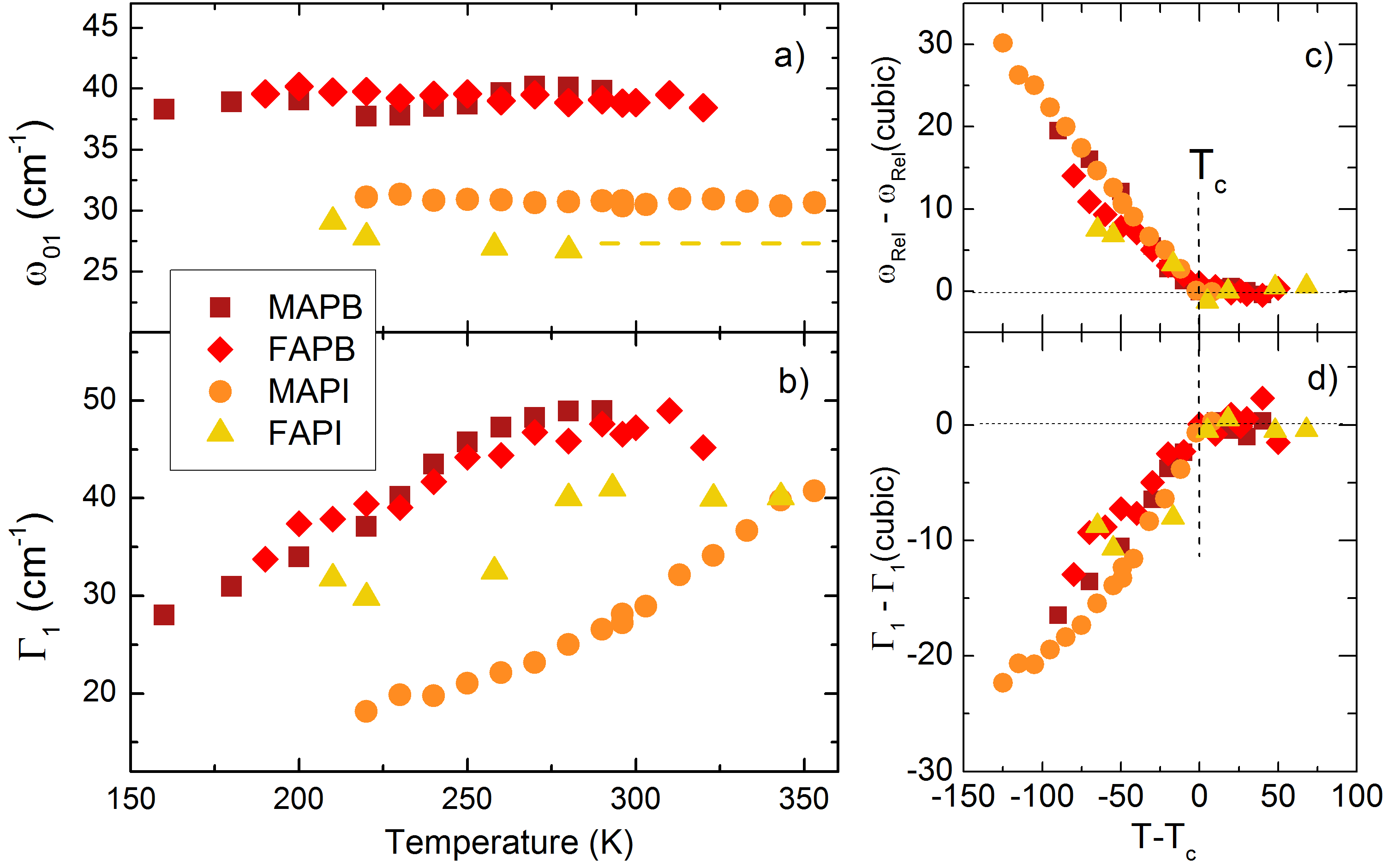}
\includegraphics[width=6.6cm]{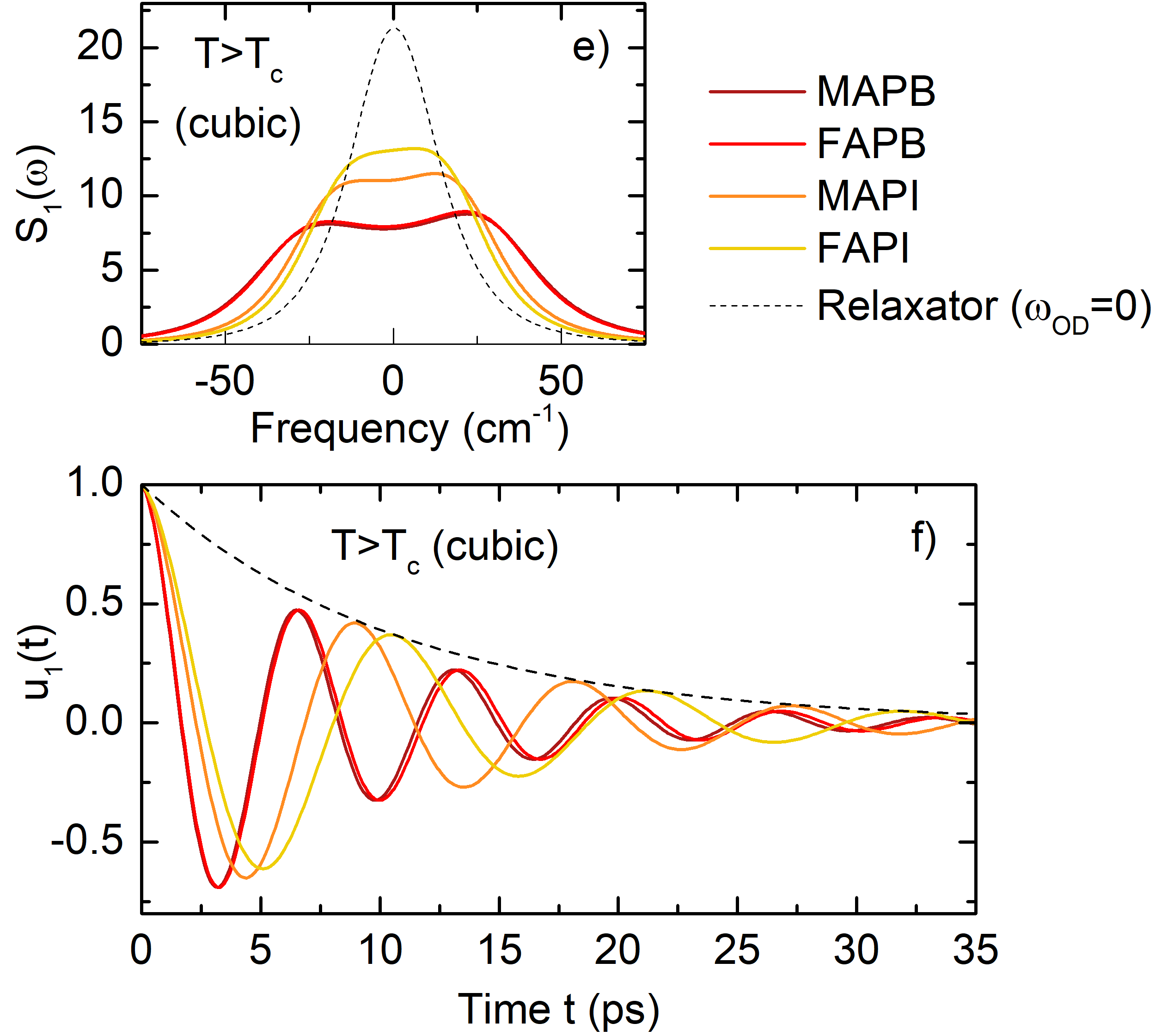}
\caption{
DHO-model: fitting parameters of the Raman mode of lowest frequency $\omega_1$ in MAPB, FAPB, MAPI, and FAPI. 
{\bf a)} Temperature dependence of the frequencies $\omega_{01}$ and {\bf b)} damping $\Gamma_1$. 
{\bf c)} Normalized relaxational frequencies $\omega_{Rel}$(T)-$\omega_{Rel}$(T$\geq$T$_c$) as a fonction of T-T$_c$, and  {\bf d)} normalized damping $\Gamma_1$(T)-$\Gamma_1$(T$\geq$T$_c$) as a function of T-T$_c$. {\bf e)} Spectral line shapes in the cubic phase and {\bf f)} temporal response of the atomic displacements.} 
\label{fig:RSResfit1}
\end{figure*}

The Brillouin spectra  also stand on an unexpectedly strong rounded background signal most likely corresponding to the top of a broader QE component similar to MAPB but indistinguishable in Fig.~\ref{fig:BSspe}.
Accordingly, the fits in Fig.~\ref{fig:BSspe} were performed using Eq. 2 for the narrow BS QE, Lorentzian functions for the longitudinal (LA) and transverse (TA) acoustic waves, and a constant for the broad QE.

\section{Analysis and Discussion}

With respect to possible signatures of critical fluctuations around the cubic to tetragonal phase transition, we will successively discuss the behavior of the optical phonon modes, then the molecular orientational (pseudospin) degrees of freedom (RS QE, INS QE, and broad BS QE), and finally the central component (INS CP and narrow BS QE).

\subsection{Phonon modes}

For optical phonons, we will first concentrate on the fitting  of the mode $\omega_1$ whose temperature evolution  is subject to discussions \cite{Yaf17,Yan20}, in particular regarding its spectral shape which could indicate a soft mode behavior.
However, as opposed to the soft-mode model, our careful analysis reveals that the  frequency of mode 1, $\omega_{01}$, is fairly constant with temperature in MAPB, FAPB, and MAPI (Fig.\,\ref{fig:RSResfit1}a).
In FAPI, the data analysis in the cubic phase was difficult and $\omega_{01}$ had to be fixed to a constant value extrapolated
from the low-T data (dashed line in Fig.\,\ref{fig:RSResfit1}a).
Conversely to the frequencies, the four spectral dampings exhibit a clear drop down upon cooling at the onset of T$_c$ (Fig.\,\ref{fig:RSResfit1}b).  
One observes however that $\Gamma_{1}>\omega_{01}$, and  it is known that  the experimental determination of $\omega_0$ and $\Gamma$ of an highly damped oscillator is quite problematic. 
Therefore we also calculated the most reliable spectral quantity in that case, {\it i.e.} the relaxational frequency $\omega_{Rel}=\omega_{01}^2/\Gamma_1$, corresponding  to the half width at half maximum of  $S_1(\omega)$ in case of overdamped modes. 
The results are shown in Fig.\,\ref{fig:RSResfit1}c as a function of T-T$_c$ and after a normalization by the value of $\omega_{Rel}$ in the cubic phase, {\it i.e.} $\simeq$~33, $\simeq$~32, $\simeq$~23, and $\simeq$~19~cm$^{-1}$, in MAPB, FAPB, MAPI, and FAPI, respectively.
The curves overlapp very nicely, emphasizing a common dynamical behavior in the four materials. 
The statistical error on $\omega_{Rel}$ is smaller than that on $\omega_{01}$ and $\Gamma_1$,  allowing thereby refining the values of the transition temperature : 
T$_c$(MAPB)\,=\,245\,K$\,\pm\,$\,10\,K, T$_c$(FAPB)\,=\,270\,K$\,\pm$\,10\,K, (it was estimated to 265 K in Ref. \cite{Fer18}),
and T$_c$(FAPI)\,=\,275\,K$\,\pm$\,10\,K.
For MAPI, we find T$_c$\,=\,345\,K a value rather close to T$_c$\,=\,335\,K$\,\pm$\,10\,K found by preliminary elastic neutron scattering experiments.
A similar normalization also works for the width (Fig.\,\ref{fig:RSResfit1}d) with however a higher uncertainty in the definition of T$_c$. 

Within the DHO model, another quantity that can be extracted for lattice modes with large dampings is 
$\omega_{OD}  =\pm\sqrt{\omega_{01}^2-\Gamma_1^2/4}$, i.e. the frequency of the temporal response $u_1(t)$ of the oscillator in the absence of driving force, 
$u_1(t)=u_{01}e^{-t\Gamma_1/2}\cos(\omega_{OD}t+\phi)$.
$\omega_{OD} = 0$ (dashed line in Fig.\,\ref{fig:RSResfit1}f) defines the overdamped limit (OD), that is the transition from a vibrational regime  when $\omega_{01}>\Gamma_1/2$ toward a relaxational one when $\omega_{01}<\Gamma_1/2$.  
Within this model, and although strongly damped, the vibration $\omega_1$ in the four compounds remains vibrational-like rather than relaxational-like (Fig.\,\ref{fig:RSResfit1}f).  
Another interesting quantity is the maximum of the phonon response $S_1(\omega)$ at $\omega_{MAX} =\pm\sqrt{\omega_{01}^2-\Gamma_1^2/2}$.
The limit $\omega_{MAX}=0 $ ({\it i.e.} $\omega_{01}=\Gamma_1\sqrt{2}$), is reached when the Stoke and anti-Stokes maxima of $S_1(\omega)$ merge into one single peak centered at $\omega$=0.
$S_1(\omega)$ are plotted in Fig.\,\ref{fig:RSResfit1}e using the fitted values of $\omega_{01}$ and $\Gamma_1$ obtained in the cubic phase and averaged over all the data. 
The responses of MAPB and FAPB are very similar and almost superpose. 
Considering the large error bars on FAPI data, the responses of  MAPI and FAPI are also very similar but stand at lower frequency than the two former. The response function of a vibration at the  crossover toward a relaxational regime ($\omega_{OD} = 0$) is shown for comparison (dashed line). 
This shows that although  overly damped, the phonon actually remain defined even in the cubic phase. 
The same conclusion can be made for the modes measured by INS at the  M- or R- points, that have been discussed above.

The renormalization of the vibrational responses into master curves  $\Gamma_1$(T-T$_c$) and $\omega_{Rel}$(T-T$_c$) (Figure \ref{fig:RSResfit1}c,d) strongly supports an unified description of the vibrational properties in the four compounds.
For example,  the frequency $\omega_{01}$ of the low frequency mode is constant when passing through the cubic-tetragonal transition (Fig.\,\ref{fig:RSResfit1}a)  and therefore  this vibration  cannot  be considered as a soft-mode. 
Moreover, the two bromide samples exhibit similar frequency and damping in the cubic phase, and same arises for the two iodide samples (Fig.\,\ref{fig:RSResfit1}a,b). 
Their frequency is in addition inversely proportional to the square root of the mass of the halide atom, {\it i.e.} $\omega_{01}\propto1/\sqrt{M}$, so that
$\omega_{01}({\rm bromines})$\,/\,$\omega_{01}({\rm iodines})=\sqrt{M_I/M_{Br}}$. 
This nicely supports the assignemnt of the mode to motion of the  PbX$_6$ halide octahedra \cite{Fer20}.
A comparison of the width and lineshape of $\omega_1$ with {\it ab initio} simulations \cite{Yaf17} suggests in addition that the mode  contains a significant part of bending motions of the PbX$_6$ cage, with a possible weak contribution of stretching at low frequency (see Fig. S2 in the supplemental material \cite{SupMat}).  
However, as pointed out above, fitting the low frequency Raman response with a broad Debye relaxation as done in Ref. \cite{Yaf17} provides a very poor agreement as compared to the DHO model.

\begin{figure}[h]
\includegraphics[width=6cm]{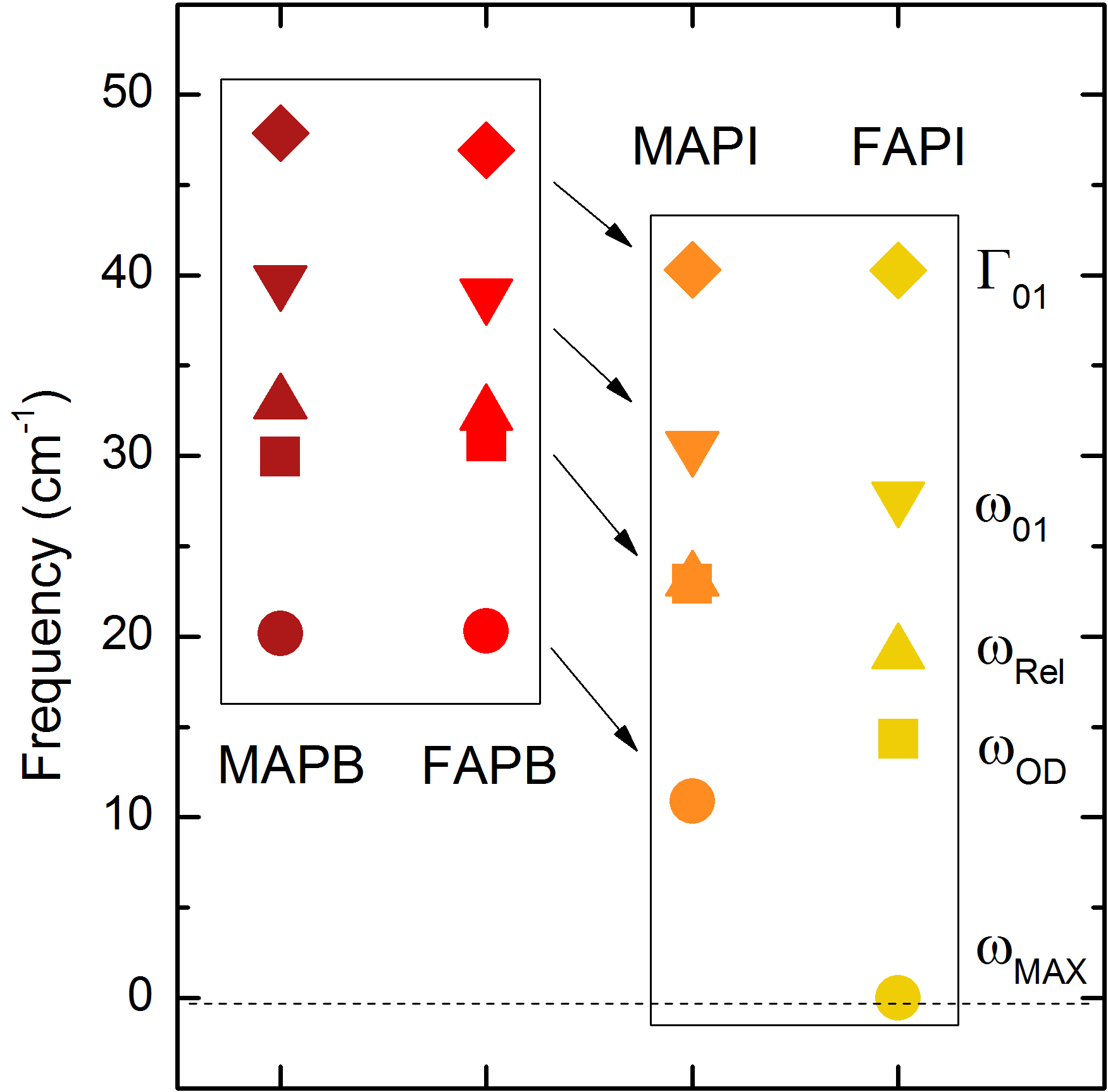}
\caption{Characteristic frequencies and linewidth of the low frequency Raman mode of MAPB, FAPB, MAPI, and FAPI, in the cubic phase.} 
\label{fig:RSResfit2}
\end{figure}

The parameters deduced from the Raman results are summarized  in Figure \ref{fig:RSResfit2}, where the linewidth $\Gamma_{01}$, the frequencies $\omega_{MAX}$, $\omega_{OD}$, $\omega_{Rel}$, $\omega_{01}$ are shown for the four compounds.
FAPI exhibits the softest behavior with regard to all its fitting parameters.  
Despite reaching the critical regime ($\omega_{MAX}=0$), the mode conserves its vibrational character, as shown by the sinusoidal shape of its temporal response in Fig.\,\ref{fig:RSResfit1}f ($\omega_{OD} > 0$). 
Therefore, it shows that the relaxational frequency associated to the lowest optical mode undergoes a universal slowing down in HOPs when approaching the phase transition from the tetragonal phase, but does not exhibit neither a  full condensation  nor a hardening in the high temperature phase  expected for a displacive phase transition. 
In the extended pseudospin-phonon framework, it is consistent with a critical dynamics attributed to a pseudospin with a very slow relaxation. At this point, we shall discuss the two possible origins of such a low and ultra-low frequency dynamics in HOPs.

\subsection{Molecular (pseudospin) motions}

Considering the dynamical behavior of the organic molecules and related possible pseudospin variables,  two main types of relaxational motions are proposed \cite{Che15,Leg15,Bak15}: i) rotations around the C-N axis, ii) whole reorientation of the C-N axis of the molecule, e.g.  between faces, corners, or edges of the pseudo cubic cage, this latter motions having a polar character. 
But other relaxations are also possible such as wobbling of the C-N axis in a cone (precession-like motion) \cite{Bak15}. 
To these reorientational motions adds slower relaxations associated to translations of the center of mass of the organic molecule \cite{Yaf17}, pointing-out the large spread of possible movements of the molecule inside its PbX$_6$ cage. 
Table S1 in the supplemental material compiles the results obtained at room temperature (or close to the cubic to tetragonal transition) using different techniques for the reorientational motions \cite{SupMat}. 
One notices that the characteristic relaxation times are typically faster than $\sim$~10~ps, leading to spectral linewidths of several wave numbers (few tenths of meV).  
The reasons for the very wide distribution of the numerical values in Table S1 are manifolds.
The large variety of motions combined to the distorted PbX$_6$ environment of the cations likely give rise to a distribution of relaxation times, for which each technique is more or less sensitive. 
Data processing can also be questioned. For example, a loose definition of the signal underneath the QE arising from the broad $\omega_1$ mode can easily generate significant differences on the fitting parameters.

\begin{figure}[h]
\includegraphics[width=8cm]{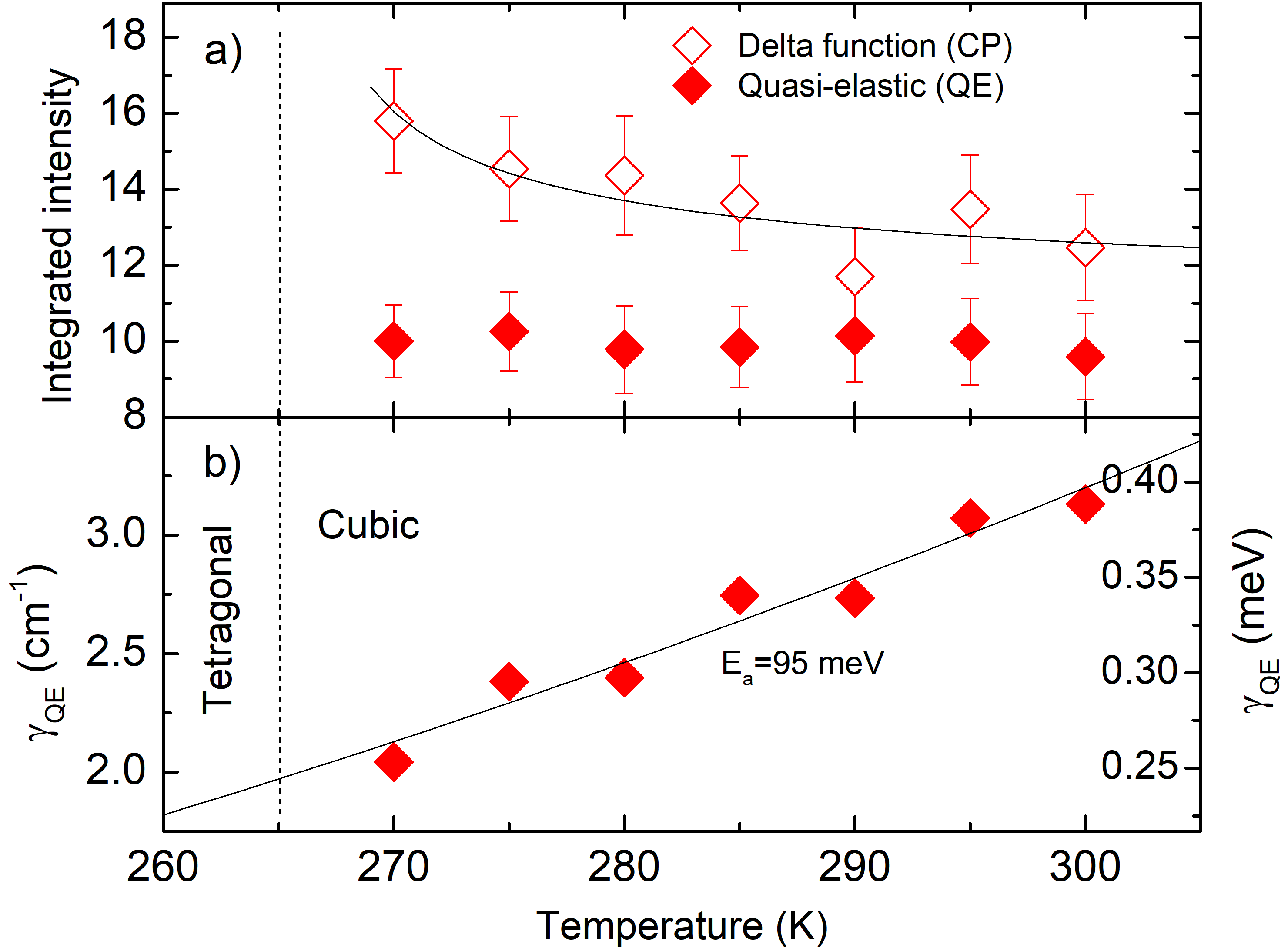} 
\caption{
Analysis of the quasi-elastic INS response in FAPB at the  M-point {\bf Q}=(3/2 1/2 0).  
{\bf a)} Area of the QE and CP components: $A_{QE}\times\gamma_{QE}$ for the former (filled squares), and the delta function amplitude for the latter in addition to the flat incoherent scattering (open squares). The line is a guide for the eyes.
{\bf b)} Temperature dependence of the damping of the QE component, $\gamma_{QE}$ fitted with an Arrhenius law (line).
}
\label{fig:QEINS2}
\end{figure}

The temperature dependence of the peak area and the width of the QE, $\gamma_{QE}$, resulting from the fits of the INS data in Fig.~\ref{fig:QEINS1} are shown in Fig.~\ref{fig:QEINS2}. 
The fit of the QE with an Arrhenius law provides a very good agreement with experiment, with in addition an activation energy $E_a\simeq 95 $~meV very close to literature data~\cite{Che15}.
Therefore, the limited reduction of the  width does not seem to be related to critical fluctuations. 
The absence of critical behavior is also supported by its structure factor, $A_{QE} \gamma_{QE}$, which remains constant within the temperature range explored (Fig.~\ref{fig:QEINS2}.a). 

The characteristic relaxation times of the RS and INS QE in this work, as well as that of the broad Brillouin QE ~\cite{Let16}, are in the range of those attributed in the literature to reorientational-type motions~\cite{Ono90,Che15,Leg15} (see also Table S1 in the supplemental material \cite{SupMat}).
For those, one common behavior is that the temperature dependence of the relaxation time follows an Arrhenius law with an activation energy affected (or not) by the transition.
We demonstrate here in addition, that these motions are not critical at the transition. 
The entropies reported  at the cubic to tetragonal phase transitions in HOPs \cite{Ono90} point toward an order-disorder behavior consistent with the freezing of these molecular pseudospins, but it is not proving that these degrees of freedom  are driving the structural phase transitions. 
In the language of the Landau theory of phase transitions, the molecular pseudospins related to C-N axis reorientations rather appear as secondary order parameters \cite{Par85, Mar91}. 
Indeed, similar cubic to tetragonal  phase transitions occur in inorganic compounds, where such reorientational motions of the cations are absent. 
These materials exhibit low-frequency modes along the R and M points but here again, a clear displacive (condensation at T$_c$)  of these modes is not reported \cite{Lan21}. 
It is therefore likely that the dynamical  behaviors controlling the phase transitions occur at lower frequencies than the tenth of meV ($\sim$ cm$^{-1}$) accessible by our INS  and Raman experiments.
This corresponds to ultra-low frequency motions as compared to those of low-lying optic-phonons.

\subsection{Ultra-low frequency motions}

Conversely to the INS QE, the delta function amplitude slightly increases upon cooling down to the phase transition (Fig.~\ref{fig:QEINS2}.a).
If it would be solely incoherent scattering, its intensity should not change much with temperature in this limited window. 
 According to the numerous reports of thermally activated behaviour, the incoherent intensity would actually even slightly decreased upon cooling following an Arrhenius law \cite{Swa15}. 
In contrast, we observe an increase of the delta function intensity which suggests a CP contribution as it has recently been discussed in MAPI as well~\cite{Wea20}. 
They both exhibit pretransitional fluctuations when approaching T$_c$,
but without substantial divergence at the transition.
This behavior points towards an order-disorder transition associated with longer characteristic times, not resolved by our experiment, similarly to many other phase transitions~\cite{Sha72,Bou96}. 

\begin{figure}[h]
\includegraphics[width=7.5cm]{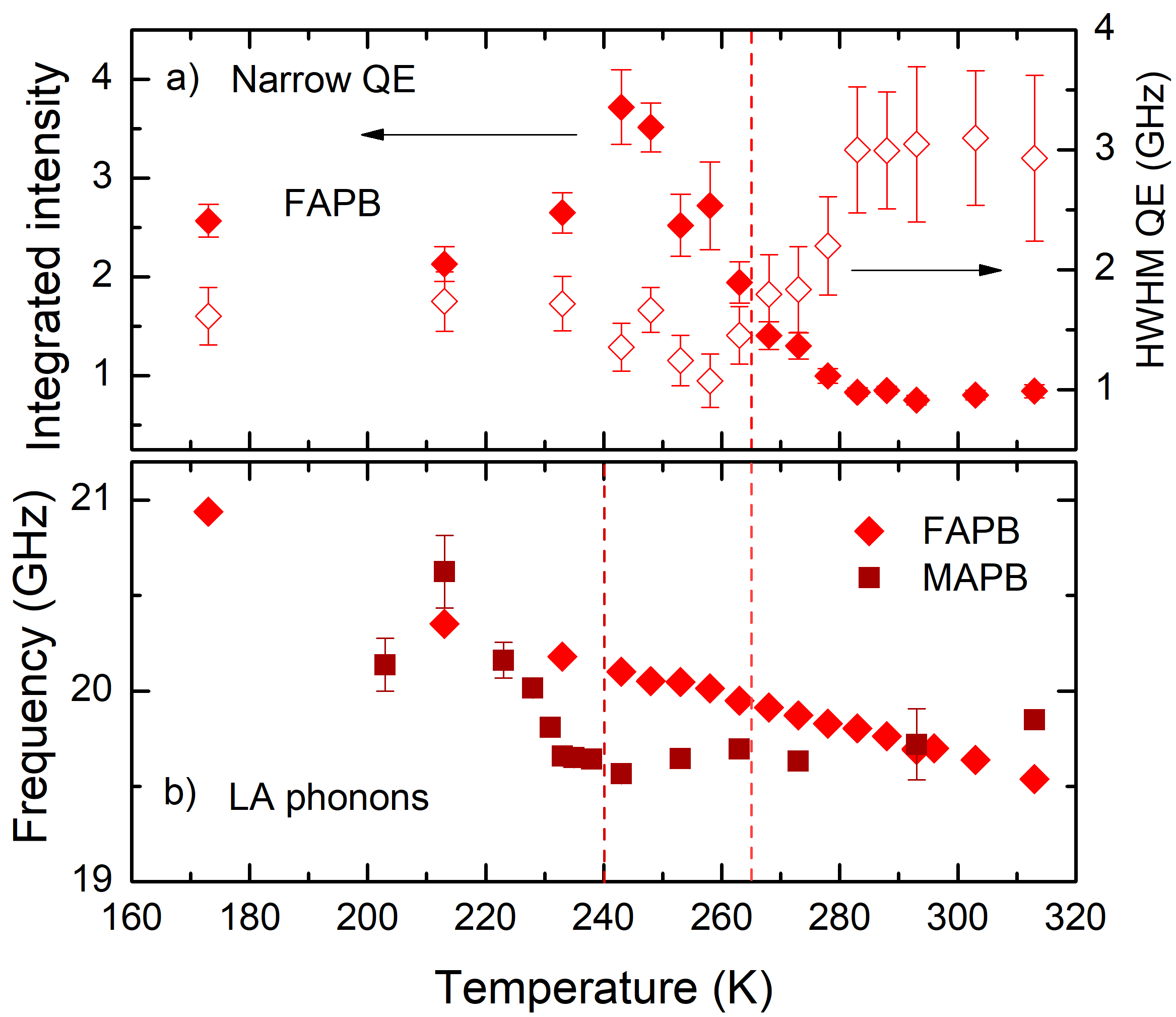}
\caption{{\bf a)} Area (full symbols) and half width at half maximum (empty symbols) of the narrow QE in FAPB, and {\bf b)} frequency of the LA phonon associated to the elastic constant C$_{11}$ in FAPB and MAPB, as a function of temperature. The dashed lines indicate the cubic to tetragonal transition in MAPB (240 K) and FAPB (265 K).} 
\label{fig:BrilFit}
\end{figure}

Looking at the ultra-low frequency of Brillouin scattering, one observes that  the narrow BS QE in FAPB also experiences a weak but visible critical-like behavior on cooling through T$_c$ (Fig.~\ref{fig:BrilFit}a).  
Its intensity reaches a maximum within the temperature range [140-160 K], namely slightly above T$_c$. 
However, the difference might be due to the local laser heating of the sample.  
The transition also induces a reduction of its linewidth, from $\sim$~3~GHz in the cubic phase to $\sim$~1.5~GHz in the tetragonal phase.  
These observations are fully consistent with our INS results if one associates the narrow  BS QE  to the (non-resolved) INS CP, and the broad BS QE to the INS QE (Fig.~\ref{fig:QEINS1}).
Likewise, we anticipate that the CP reported in MAPI~\cite{Wea20} is also dynamical with a characteristic inverse relaxation time in the GHz range.
Unfortunately, the weakness of the observed critical behaviors prevents from extracting critical exponents as usually done for phase transitions.
The activity of these critical excitations in BS at zone center, while expected at the M-point where occurs the tetragonal instability, is likely a further indication of a strong molecular disorder. 
In these materials, the disorder gives rise to an incoherent scattering channel in light scattering due to the partial meaningless of the wave vector $q$ that lifts the momentum selection rules. 

The weak enhancement of the BS QE and INS CP could suggest a limited growing of quasi-static clusters with the tetragonal symmetry above the transition, possibly nucleated around defects \cite{Hal76}.
It could be halogen (particularly iodine) vacancies which are the most harmful for optoelectronic properties and for devices under operation, but we cannot exclude in the present case that more classical extended defects, such as grain boundaries, stacking faults or dislocations, might also contribute.
These observations 
highlights an underlying universal behavior in HOPs connected to the central peak component. 
The strong anharmonic motion of the cation center of mass  in inorganic perovskites \cite{Yaf17} is the only remaining degree of freedom sharing similarities to the rattling of the cations in the inorganic cages \cite{Fer20}, and undergoing a possible critical behavior at ultra-low frequency. As such it could correspond to the primary order parameter. 
The importance of defects was put forward for MAPI as a possible mechanism at the origin of both a slightly first order character of the transition, and the incomplete divergence of the central peak. We cannot confirm the existence of a slightly first order character for the transition in FAPB, but  both a displacive mechanism and a critical behavior of the reorientational motions of the organic molecules can be ruled out from our combined Raman and neutron scattering study for the four HOPs.
The existence of a central peak is the only scenario able to reconcile all the observations on both HOPs and inorganic perovskites. 

Finally, It is worth noting  that contrary to the lowest frequency optical mode $\omega_1$ which behave exactly the same way in the four HOPs at the cubic to tetragonal transition (Fig.~\ref{fig:RSResfit1}), the temperature dependence of the LA phonon associated to the elastic constant C$_{11}$ in MAPB and FAPB is quite different, at least at Brillouin frequencies.
In the former compound the longitudinal mode is affected by the transition and couples to the narrow QE~\cite{Let16}, while not in the latter (Fig.~9b).
This is attributed to the different  tetragonal crystalline symmetries, $I4/mcm$  and $P4/mbm$, respectively.
The strain imposed by a transition towards a structure having the $I4/mcm$-symmetry must affects the LA phonons~\cite{Let16, Guo17}.  
Conversely no effects are expected on LA phonons for a transition towards $P4/mbm$ since in that case the transition does not generate experimentally strain tensor components~\cite{Sch17, Car07}.

From the theoretical viewpoint, the overwhelming anharmonicity of the perovskite lattice at high temperature  has been proposed to induce both low frequency phonons at the R and M edges of the Brillouin zone \cite{Yaf17, Mar17} and stochastic ionic motions related to fluctuations at the $\Gamma$ point strongly affecting the carrier mobilities \cite{Mar18, Kat18}. 
We shall point out a slight difference between these two cases: in the former case, the off-center disordered motion of the cation, tentatively assigned here to the narrow INS CP and possibly the primary order-parameter, is decoupled from the halogen motions. 
In the latter case the cation motion is coupled to polar fluctuations involving all the atoms.

\section{Conclusion} 
 
No slowing down of a low-frequency soft vibration was observed in HOP single crystals, neither at the Brillouin zone center (Raman) nor at the zone boundary close to superstructure peaks (INS). 
This appears to be a specificity of lead halide perovskites as compared to other perovskites compounds.  
These observations strongly suggest the critical importance of relaxational motions, very similar to the case of plastic crystals. 
However, the reorientational motions of the molecular cations have not evidenced any critical behavior despite their underlying order-disorder character. 
This is likely a marker of very weak intermolecular interactions and of a secondary order-parameter in the framework of the Landau theory of phase transitions.
Therefore, the general scenario describing the very  similar structural phase transitions in both inorganic and hybrid halide perovskites, is rather related to the strongly anharmonic and ultra-low frequency motions of the cation center of masses in the perovskite cages, which might be at the origin of the narrow QE observed in Brillouin scattering and leading to the not-resolved central peak in INS.  
However, the present observation by neutron scattering in FAPB, consistent with previously reported results for MAPI, further points toward the general importance of defects in HOPs preventing the net divergence of the order parameter correlation lengths at the critical temperatures.

\subsection{Aknowledgments}

J.E. acknowledges support from Institut Universitaire de France. C.K. acknowledges support from Agence Nationale pour la Recherche (MORELESS project). 



\end{document}